# Hydrothermal synthesis of SnO$_2$ particles for the degradation of Methylene Blue (MB) dye in presence of sunlight

Komal Singh, Jyoti Sindkar, Mrunal Ramdasi, Vrishali Jadhav, Rohidas B. Kale[*]

Department of Physics, The Institute of Science, Dr. Homi Bhabha State University, Madam Cama Road, Mumbai, India

ABSTRACT

Three distinct samples were proceed through synthesis utilizing the hydrothermal method to develop tin dioxide (SnO$_2$) nanoparticles. Consistency in all other parameters was ensured by maintaining a constant temperature and time throughout the synthesis. X-ray diffraction (XRD) and scanning electron microscopy (SEM) were used to examine how the surfactant affected the structural, morphological, and crystallographic characteristics. A tetragonal rutile SnO$_2$ phase was confirmed to have formed by XRD investigation, with high crystallinity indicated by strong diffraction peaks. Higher precursor concentration samples showed aggregation, indicating that the smaller size of the nanoparticles caused them to interact. These findings show that changing the surfactant has a major impact on the crystallinity, size, and shape of SnO$_2$ nanoparticles, which makes this technique establish for certain uses including energy storage, gas sensors, and photocatalysis.

Additionally, we examined the photocatalytic activity of the aforementioned samples, indicating from the UV-Vis characterization results that the sample containing PEG as a surfactant performs better than others

## 1.Introduction

Water is the lifeblood of all known forms of life.[1] From hydration, an essential biological function, to acting as a solvent, coolant, and transport medium, water's roles are multifaceted and indispensable [2-4]. For humans, animals, aquatic life, and plants, water ensures not only survival but also the maintenance of ecological balance and biodiversity. [5-7]

Now-a-days, most of the synthetic dyes and pigments are toxic nonetheless, they are utilised in the biomedical field, textile, foodstuff, paper, leather and printing industries. The discharge of waste water from these industries into natural streams that leads to severe problems not only to human beings but also to the environment. This contaminated water demands high chemical oxygen and biochemical oxygen since the dye colours block the sunlight access to aquatic life and reduces the action of photosynthesis with the ecosystem [8]. Among the organic contaminants, chemical dyes have become an indispensable concern for researchers due to their toxic and carcinogenic nature [9,10]. The dyes are used extensively in textiles, leather, food, agriculture, photoelectrochemical cells, cosmetics, and wastewater [11-13].

The refinement of waste water is one of the biggest environmental challenges of the present day. A lot of physical and chemical processes such as coagulation, membrane filtration, ozonation, oxidation, bio-degradation etc have been broadly used for the cleaning of dye-containing waste water.[14]. However, these processes for advanced water treatment have been limited due to the high cost for large-scale applications, low water treatment volume, and high chemical usage [15-18].

In recent years, photocatalytic processes have been thoroughly examined as a green and safe technology for water purification and dye degradation because of their high efficiency, non-toxicity, low cost, and capability to disinfect/degrade a broad range of microorganisms and dye under the sun lights containing UV rays. [19-21]. During the photocatalysis, the source of light and catalysts are used to accelerate the process of the degradation of organic pollutants. According to the source data, photocatalysis can remove 70−80% of pigments from industrial effluent [22].

In the course of photocatalytic degradation, NPs are activated in the presence of UV/Vis radiation which in-turn creates a redox environment in the system and behaves as a sensitizer for light-induced redox mechanisms [23]. The available literature shows the degradation of different dyes in the effluents under UV-Vis radiation using photoreactor [24-30] as shown in Table 1. Since the efficiency of dye degradation has increased in the presence of nanomaterials as photocatalyst [31-33], there is a demand for new strategies to synthesize different nanomaterials with desired physical-chemical properties and optical-electronic properties. Nanomaterials with improved properties demonstrate versatile applications in catalysis, gas sensors, adsorbents, energy storage devices and biological applications [34-39]. Hitherto different types of metal and metal oxide NP are synthesized for various applications, among them $SnO_2$ is one of the prominent and well investigated materials with different applications in various fields like catalysis, gas sensors, energy storage devices, dye based solar cells, transparent conducting electrodes and biomedical applications [26,39-44]. $SnO_2$ can be utilized as a captivating n-type semiconductor, since its band gap is 3.6 eV with 130 eV for excitation binding energy. Moreover, the absorption of UV light for this material is found to be promising and similar to that of $TiO_2$ [45-47]

$SnO_2$ shows a tetragonal rutile structure (lattice constant : a = b= 4.7374 A°, c = 3.1864 A°.) Unit cell consist of 2-six fold coordinated Sn atom ($Sn^{4+}$) and 4 three - fold coordinated O atoms ($O^{2-}$) [48-50]. The crystalline $SnO_2$ solid or $SnO_2$ powder is white or off- white in colour. Boiling point and melting point of $SnO_2$ is 1800-1900°c and 1500-1600°c, respectively. Also $SnO_2$ is soluble in $H_2SO_4$(conc ) and HCl though it is water- insoluble [51-52].

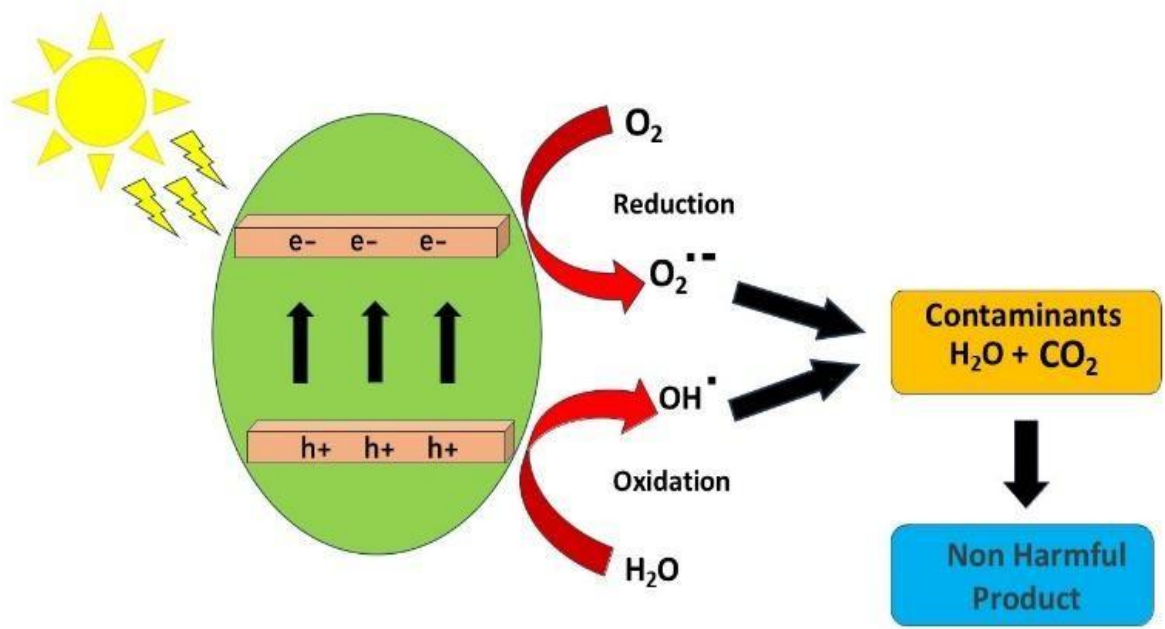

**Fig. 1**

When $SnO_2$ is photoinduced by a photon of energy 'hv′ greater than or equal to the excitation energy 'Eg', electrons from the valence band jump to the empty conduction band that creates e− - h+ pairs. As the e⁻ - h⁺ pairs start moving toward the surface of $SnO_2$, which is essential in redox reaction. Hydroxyl radicals are obtained from the reaction of $H^+$ with water, while hydroxide ions and superoxide radicals are created due to the reaction of electrons with oxygen. The produced superoxide radical anions further react with $H^+$ that led to the generation of hydrogen peroxide. The reaction of hydrogen peroxide with superoxide radicals could pave the way for the formation of hydroxyl radicals. The obtained hydroxyl radicals are found to be excellent oxidizing agents such that they can attack the organic pollutants whereby they are adsorbed on the surface of $SnO_2$ resulting in the rapid production of intermediate compounds. The resulting intermediate compounds will then be converted to green products like $CO_2$, $H_2O$ and mineral acids [53].

| Material | Method of Synthesis | Experimental of Synthesis | Light Source | Time of Degradation | Percentage | Reference |
|---|---|---|---|---|---|---|
| $SnO_2$ rods | Solvothermal ($150^0$/12 hr) | P= 20 mg/L, 100 ml, C = 10 mg | 250W mercury lamp | 90 min | 96.30% | [63] |
| $SnO_2$ hollow microsphere | Hydrothermal (190 °C/24 hr) | P= 10 mg/L, 200 ml, C = 0.1 gm | UV irradiation using 500W Hg lamp | 25 min | 95% | [64] |
| $SnO_2$ nanoparticles | Hydrothermal ($135^0$/24 hr) | P= 10 mg/L, 30 ml, C = 00.3 mg | UV illumination | 30 min | 88.88% | [65] |
| | | | | 120 min | 90% | |
| $SnO_2$ nanoparticles | Hydrothermal ($85^0$/3 hr) | P= 75 ml, C= 50 mg | UV light | 50 min | 100% | [66] |
| $SnO_2$ nanoparticles | Hydrothermal | C=500 mg, P=8/5 mg/L, 500 ml | UV light | 180 min | 81% | [67] |
| $SnO_2$ nanoparticles | Hydrothermal ($160^0$C/12hr) | P=200ml, C=5mg | sunlight | 120 min | 80% | [68] |

**Table No. 1**

In recent years, photocatalytic technologies have demonstrated great potential for mitigating energy shortages and environmental pollutants [54,55] . Visible-light-driven photocatalysis has the noteworthy advantage of efficiently harnessing the vast energy content of solar radiation as a clean, economical, and sustainable energy source. Visible-light-active photocatalysts have attracted a lot of attention due to their ease of production and recycling through fundamental chemical processes. The amount of light that these photocatalysts can absorb is determined by their band gap energy [54]. Photocatalytic technology and visible-light photo-redox catalysis are currently considered to be the most important approaches to solving the global energy and environmental issues. Inspired by photosynthesis in plants, synthetic organic chemists are constantly diligent to develop new chemical reactions that can be conducted in sustainable, green, and environmentally safe environments. [56, 57]. Particle size, shape, composition, structure, and flaws mainly affects the physicochemical characteristics of nanomaterials, In addition processing conditions have significant impact on these factors. There are two processes generally used to synthesize np: Top-down method in which np are created by breaking down bulk materials. However in bottom-up method assembles np's beyond a wide scale through atoms or molecules. For exact control over material properties, solution-based bottom-up synthesis is frequently employed. Among these solution-based techniques, hydrothermal synthesis is particularly well-liked due to its affordability, ease of use, efficiency, and affordability in producing a wide range of sophisticated materials (organic, inorganic, or hybrid) with precision in morphologies and regulated particle sizes [58].

## 2. Experimental section

2.1 Chemicals used

Stannous chloride ($SnCl_2.2H_2O$), Sodium hydroxide (NaOH), ethanol, Cetyltrimethylammonium Bromide (CTAB), Polyethylene glycol (PEG).

2.2 Sample Preparation

S1 ($SnO_2$ +NaOH)

0.02M of $SnCl_2.2H_2O$ was fully dissolved in distilled water and ethanol in a volume ratio of 1:1 in order to create $SnO_2$ particles. Additionally, 10 M NaOH was added to the solution, and after 10 minutes of stirring, the mixture was transferred. to a stainless-steel autoclave lined with Teflon. Then, an autoclave was held for 12 h at 180°C in a hot air oven and finally air-cooled to room temperature. The final product was collected, washed with distilled water several times. This solution is dried for 5–6 h at 100°C and a white powder of $SnO_2$ was obtained, further used for characterization and named as $S_1$.

S2 and S3 ($SnO_2$ +CTAB & $SnO_2$ +PEG)

In order to create $SnO_2$ particles, 0.02M of stannous chloride ($SnCl_2.2H_2O$) was fully dissolved in 250ml of distilled water and ethanol in a 1:1 ratio.
Another solution of 0.01 M was prepared by dissolving CTAB into distilled water. At the same time Sodium hydroxide (NaOH) solution of 10 M was prepared.
0.01M of CTAB solution was added into the stannous chloride + ethanol mixture after dissolved all the particles.
Then a NaOH solution was added to maintain the pH as above.
A partially transparent solution was produced after ten minutes of stirring. Additionally, the final solution was transferred to a Teflon-lined stainless steel autoclave. After 12 hours at 180°C in a hot air oven, the autoclave was allowed to air cool to ambient temperature. The finished product was obtained and repeatedly cleaned with distilled water. The solution was kept for 24 h to settle down. This solution produced a $SnO_2$ powder that was used for further characterization after drying for 5–6 hours at 100°C.

The same process which is introduced in S2 was carried out for synthesis of SnO2 particles, but the only difference was CTAB replaced by PEG.

## 3. Results and discussion

3.1 Structural analysis

The XRD pattern of the SnO₂ sample is shown in fig.2
It reveals that all the peaks of SnO2 sample are well matched with the standard JCPDS card no 00-001-0657. From generated broad peaks of SnO₂, the XRD pattern indicates the nanocrystalline nature of the SnO₂ sample. The diffraction peaks records at $2\theta = 26.66°, 38.10°, 52.23°, 64.67°, 79.09$ Correspond to (110), (200), (211), (112), (321) planes.
It is similar to the SnO₂ tetragonal configuration. No other peaks were detected in the XRD pattern confirming the sample purity.
The average crystallite size (D) of the synthesized SnO₂ was calculated using the Scherrer equation, which relates the crystallite size to the broadening of the XRD peaks. The equation is: $D = k\lambda/\beta\cos\theta$
where:
D is the average crystallite size, k is shape factor (0.9), $\lambda$ is wavelength of X-ray, $\beta$ is FWHM, $\theta$ is Bragg angle [59-60].
The calculated crystalline size for NaOH, CTAB and PEG was **9.63nm, 10.53nm, 8.21nm** respectively

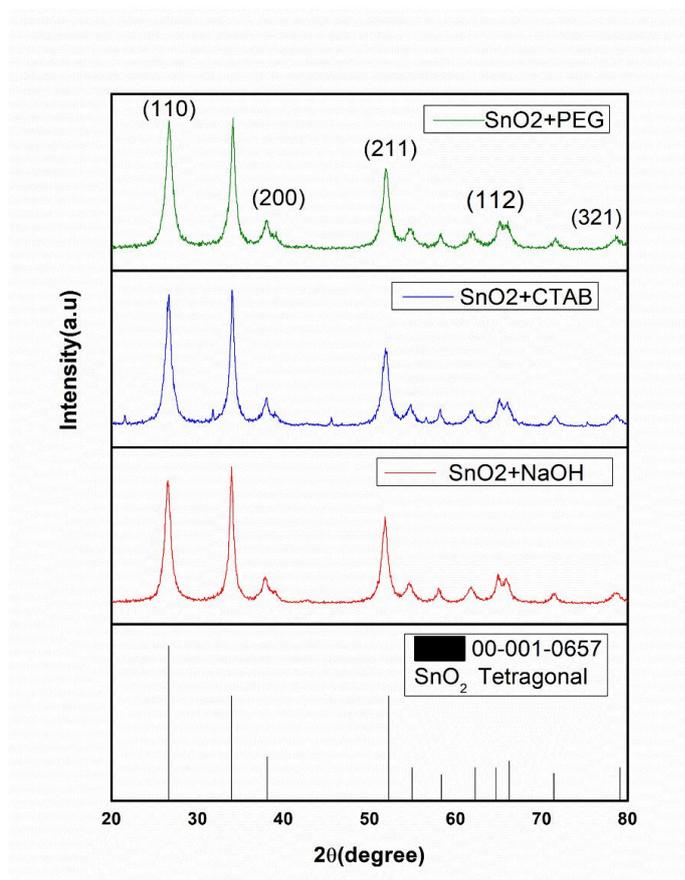

**Fig. 2**

3.2 Morphological study
The SEM images of the prepared samples, S1, S2, S3 are given in Fig. 3(a–c), respectively. It can be observed that the samples have agglomerated structure, which varies in size with the addition of surfactant. The S2 with CTAB shows a mixed morphology of some rod and agglomeration. It is evident that change in surfactant change the morphology.

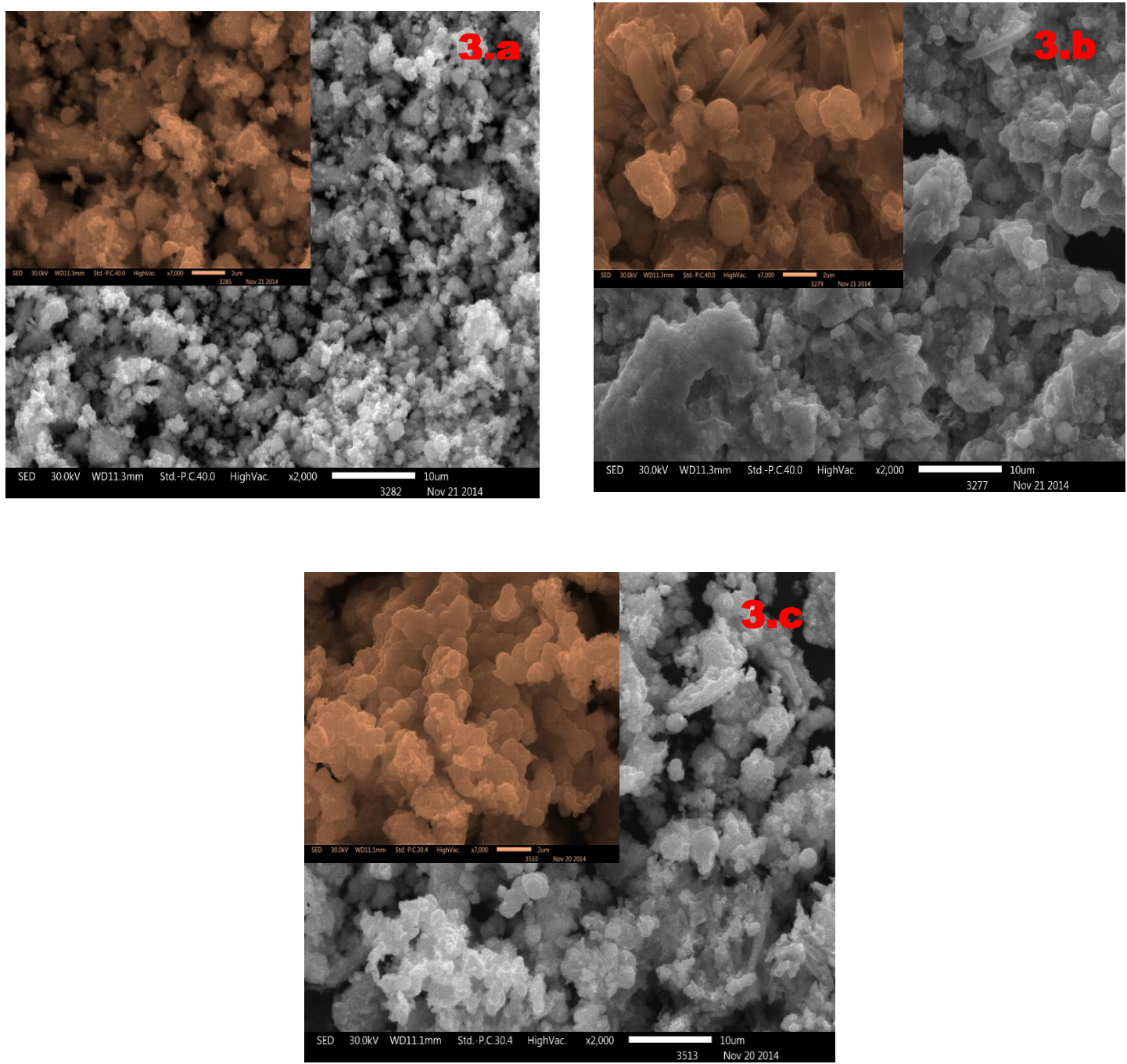

**Fig. 3**

## 3.3 UV-Vis analysis

UV–vis absorbance spectroscopy is used to monitor the optical absorption of the prepared SnO₂ NPs. As shown in Fig., the synthesized SnO₂ was observed to respond to the absorption in UV and visible regions of the wavelengths. From the UV–vis absorption spectra, band gap energy was calculated using the Tauc equation : $\alpha(\nu)h\nu = A(h\nu - E_g)^n$.

where "$E_g$" is semiconductor band gap energy, "$h\nu$" is incident photon energy, "A" is proportionality constant and "$\alpha(\nu)$" is the absorption coefficient. The exponent "n" has values of ½ and 2 for the direct and indirect allowed transitions, respectively. Fig.4(a-c) shows the graph of $h\nu$ vs $(\alpha h\nu)^2$ for S1 S2 S3 respectively an extrapolation of the Tauc plot to the x-axis gives the band gap energy. [61-62]:

The band gap energies ($E_g$) of S1, S2 and S3 are **3.6, 5.7, 3.78eV** respectively as depicted in Figure. 4

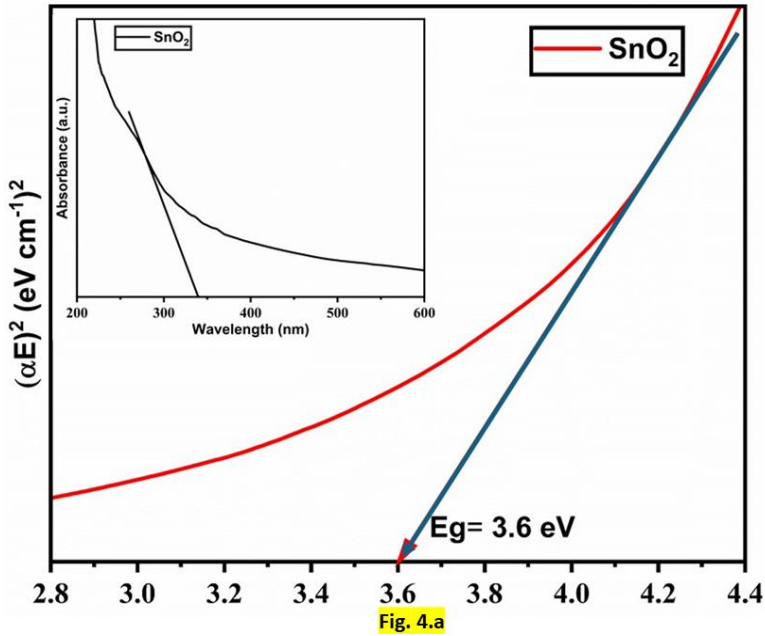

Fig. 4.a

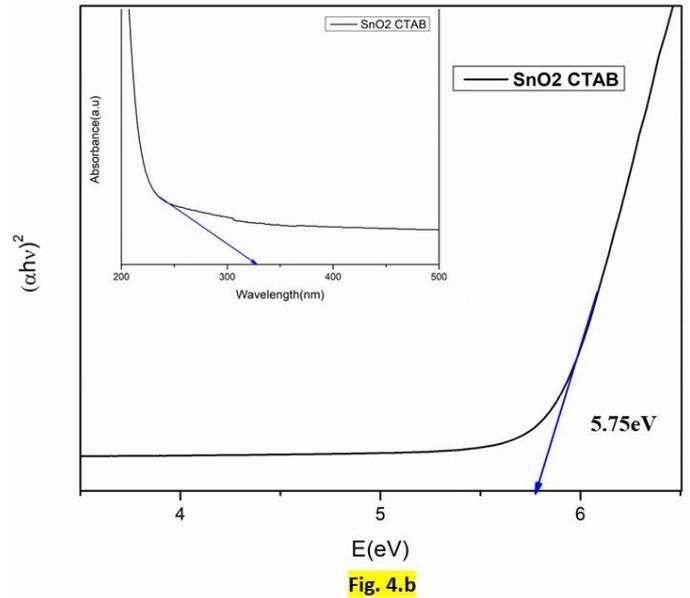

Fig. 4.b

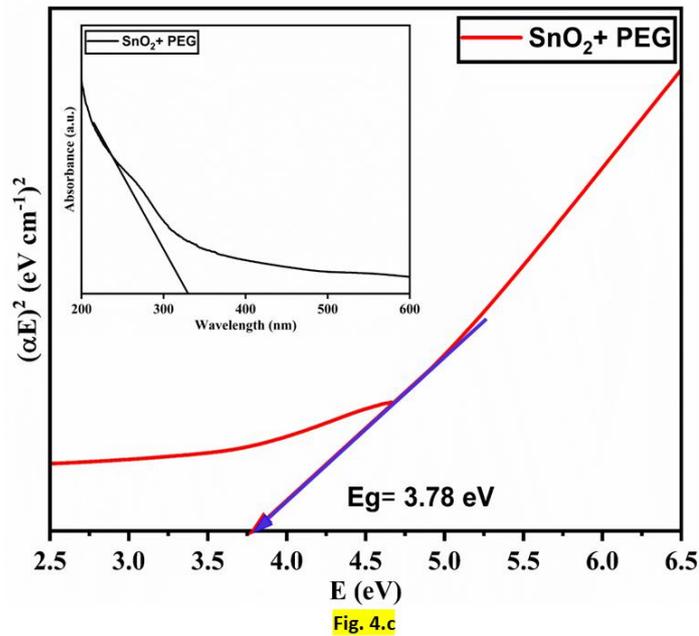

Fig. 4.c

**Fig. 4**

3.4 Photocatalytic activity

From the optical properties analysed by UV-vis S2 shows large bandgap we have selected S1 and S3 for dye degradation. The photocatalytic degradation of S1 and S3 over the time are shown in the fig. 5 and 6 respectively. For S3 the degradation is ~80% in 2.5 hours.

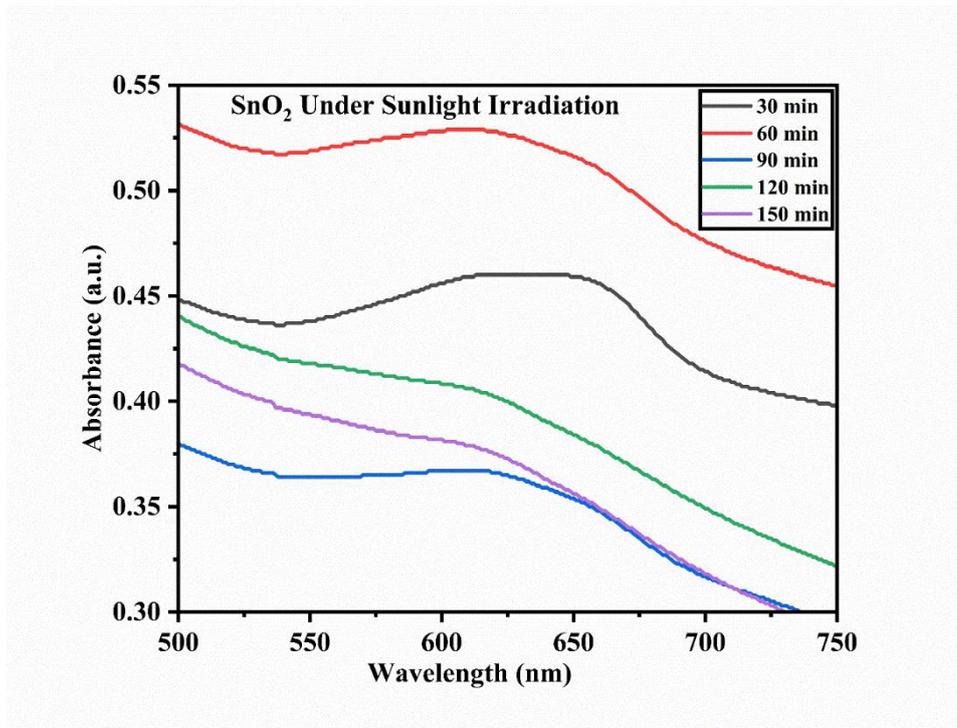

**Fig. 5**

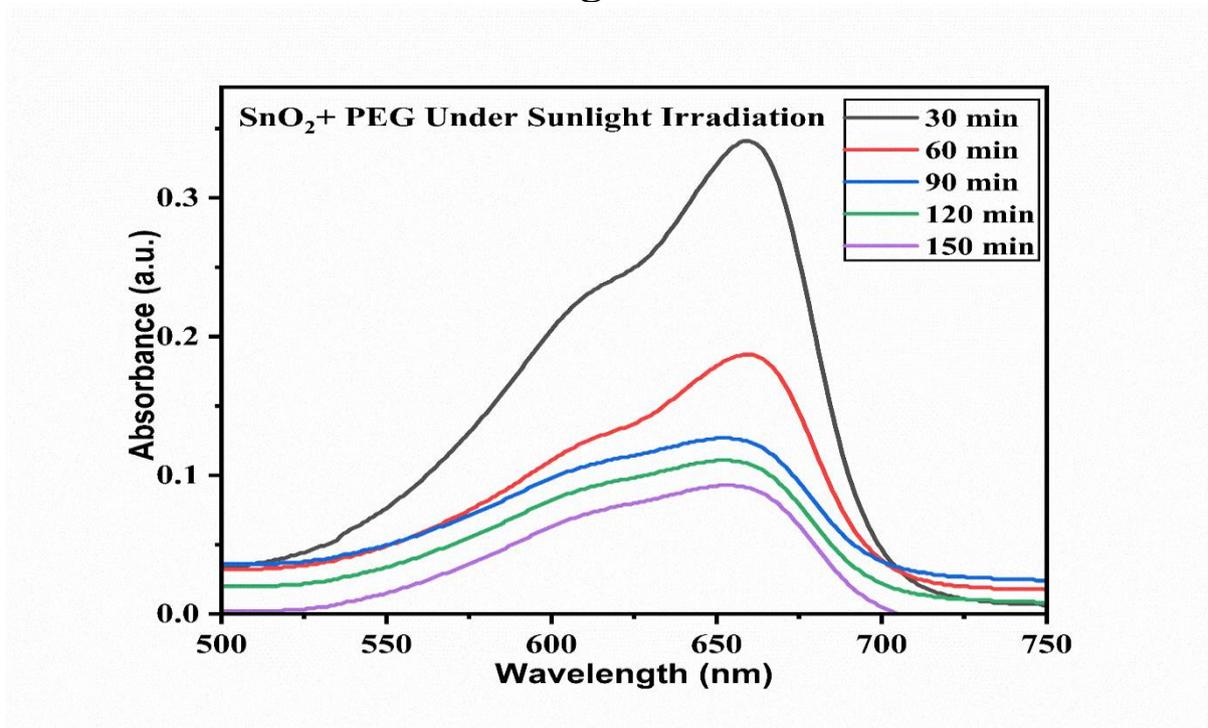

**Fig. 6**

# 4. Conclusions

In order to examine the effects of several surfactants—NaOH, CTAB, and PEG—on structural, morphological, and photocatalytic characteristics, $SnO_2$ nanoparticles were effectively created hydrothermally in this work. SEM photos showed significant morphological variations caused by the kind of surfactant, while XRD examination verified the creation of pure, tetragonal rutile-phase $SnO_2$ with nanocrystalline characteristics. The PEG-assisted $SnO_2$ sample had the best photocatalytic performance, degrading Methylene Blue by around 80% in 2.5 hours
 in the presence of sunshine, the lowest crystallite size (~8.21 nm), and a comparatively high bandgap (~3.78 eV) among the samples.

These findings show that adjusting the surfactant selection is essential for adjusting the photocatalytic effectiveness of $SnO_2$ nanoparticles. One intriguing method for creating effective, sunlight-driven photocatalysts for the breakdown of organic contaminants in wastewater is the PEG-assisted synthesis pathway. To further improve its visible-light activity and reusability in actual environmental settings, future research may investigate doping or heterostructure creation with $SnO_2$.

**References**
1. Attar, R. M., Alkhamis, K. M., Alsharief, H. H., Alaysuy, O., Alrashdi, K. S., Mattar, H., ... & El-Metwaly, N. M. (2024). Remarkable photodegradation breakdown cost, antimicrobial activity, photocatalytic efficiency, and recycling of SnO2 quantum dots throughout industrial hazardous pollutants treatment. Ceramics International, 50(19),
2. Şenol, Z. M., El Messaoudi, N., Ciğeroglu, Z., Miyah, Y., Arslanoğlu, H., Baglam, N., ... & Georgin, J. (2024). Removal of food dyes using biological materials via adsorption: a review. Food Chemistry, 139398.
3. Liu, X., & Wang, J. (2024). Decolorization and degradation of various dyes and dye-containing wastewater treatment by electron beam radiation technology: An overview. Chemosphere, 351, 141255.
4. Sirajudheen, P., Poovathumkuzhi, N. C., Vigneshwaran, S., Chelaveettil, B. M., & Meenakshi, S. (2021). Applications of chitin and chitosan based biomaterials for the adsorptive removal of textile dyes from water—A comprehensive review. Carbohydrate polymers, 273, 118604.
5. Fernandes, M., Fernandes, R. D., Padrão, J., Melro, L., Alves, C., Rodrigues, R., ... & Zille, A. (2024). Plasma in textile wastewater treatment. In Advances in Plasma Treatment of Textile Surfaces (pp. 267-322). Woodhead Publishing.
6. Sebastian, S. L., Kalivel, P., Subbiah, K., Murphy, M. A., David, J. J., & Palanichamy, J. (2024). Assessing titanium vs. aluminium electrodes for wastewater remediation in the small-scale industries (SSI) textile sector. Environmental Nanotechnology, Monitoring & Management, 21, 100950.
7. Kumar, D., Pandit, P. D., Patel, Z., Bhairappanavar, S. B., & Das, J. (2019). Perspectives, scope, advancements, and challenges of microbial technologies treating textile industry effluents. In Microbial wastewater treatment (pp. 237-260). Elsevier.
8. Suthakaran, S., Dhanapandian, S., Krishnakumar, N., & Ponpandian, N. (2019). Hydrothermal synthesis of SnO2 nanoparticles and its photocatalytic degradation of methyl violet and electrochemical performance. Materials Research Express, 6(8), 0850i3.
9. N. Jahan et al.A comparative study on the sorption behavior of graphene oxide and reduced graphene oxide towards methylene blue Case Stud. Chem. Environ. Eng.(2022)
10. Jafari Harandi, Z., Ghanavati Nasab, S., & Teimouri, A. (2019). Synthesis and characterisation of magnetic activated carbon/diopside nanocomposite for removal of reactive dyes from aqueous solutions: experimental design and optimisation. International Journal of Environmental Analytical Chemistry, 99(6), 568-594.
11. E. Brillas et al.Decontamination of wastewaters containing synthetic organic dyes by

electrochemical methods. An updated review Appl. Catal. B Environ.(2015)
12. T.A. Saleh et al.Synthesis of silica nanoparticles grafted with copolymer of acrylic acrylamide for ultra-removal of methylene blue from aquatic solutions Eur. Polym. J.(2020)
13. S.G. Nasab et al.Prediction of viscosity index and pour point in ester lubricants using quantitative structure-property relationship (QSPR)Chemom. Intell. Lab. Syst.(2018)
14. Suthakaran, S., Dhanapandian, S., Krishnakumar, N., & Ponpandian, N. (2019). Hydrothermal synthesis of SnO2 nanoparticles and its photocatalytic degradation of methyl violet and electrochemical performance. Materials Research Express, 6(8), 0850i3.
15. Zhou, H., & Smith, D. W. (2001). Advanced technologies in water and wastewater treatment. Canadian Journal of Civil Engineering, 28(S1), 49-66.
16. Teimouri, A., Nasab, S. G., Habibollahi, S., Fazel-Najafabadi, M., & Chermahini, A. N. (2015). Synthesis and characterization of a chitosan/montmorillonite/ZrO 2 nanocomposite and its application as an adsorbent for removal of fluoride. Rsc Advances, 5(9), 6771-6781.
17. Ghanavati Nasab, S., Teimouri, A., Hemmasi, M., Jafari Harandi, Z., & Javaheran Yazd, M. (2021). Removal of Cd (II) ions from aqueous solutions by nanodiopside as a novel and green adsorbent: Optimisation by response surface methodology. International Journal of Environmental Analytical Chemistry, 101(14), 2128-2149.
18. Nasab, S. G., Yazd, M. J., Marini, F., Nescatelli, R., & Biancolillo, A. (2020). Classification of honey applying high performance liquid chromatography, near-infrared spectroscopy and chemometrics. Chemometrics and Intelligent Laboratory Systems, 202, 104037.
19. Daneshvar, N., Salari, D., & Khataee, A. R. (2004). Photocatalytic degradation of azo dye acid red 14 in water on ZnO as an alternative catalyst to TiO2. Journal of photochemistry and photobiology A: chemistry, 162(2-3), 317-322.
20. Zhang, L., Yin, L., Wang, C., Lun, N., & Qi, Y. (2010). Sol− Gel Growth of hexagonal faceted ZnO prism quantum dots with polar surfaces for enhanced photocatalytic activity. ACS applied materials & interfaces, 2(6), 1769-1773.
21. Roy, H., Rahman, T. U., Khan, M. A. J. R., Al-Mamun, M. R., Islam, S. Z., Khaleque, M. A., ... & Awual, M. R. (2023). Toxic dye removal, remediation, and mechanism with doped SnO2-based nanocomposite photocatalysts: A critical review. Journal of Water Process Engineering, 54, 104069.
22. Harrelkas, F., Paulo, A., Alves, M. M., El Khadir, L., Zahraa, O., Pons, M. N., & Van Der Zee, F. P. (2008). Photocatalytic and combined anaerobic–photocatalytic treatment of textile dyes. Chemosphere, 72(11), 1816-1822.
23. D.Beydoun ,R.Amal, G.Low, S.McEvoy, Role of nanoparticles in photocatalysis, J. Nanopart. Res.1 (1999) 439-458.
24. Kowsari, E., & Ghezelbash, M. R. (2012). Ionic liquid-assisted, facile synthesis of ZnO/SnO2 nanocomposites, and investigation of their photocatalytic activity. Materials Letters, 68, 17-20.
25. Chen, W., Ghosh, D., & Chen, S. (2008). Large-scale electrochemical synthesis of SnO2 nanoparticles. journal of Materials Science, 43(15), 5291-5299.
26. Tammina, S. K., & Mandal, B. K. (2016). Tyrosine mediated synthesis of SnO2 nanoparticles and their photocatalytic activity towards Violet 4 BSN dye. Journal of Molecular Liquids, 221, 415-421.
27. Senobari, S., & Nezamzadeh-Ejhieh, A. (2018). A comprehensive study on the photocatalytic activity of coupled copper oxide-cadmium sulfide nanoparticles. Spectrochimica Acta Part A: Molecular and Biomolecular Spectroscopy, 196, 334-343.
28. M. Karimi- Shamsabadia, M. Behpour, A. Babaheidarib, Z. Saberib, Efficiently enhancing photocatalytic Activity of NiO-ZnO doped onto nanozeoliteX by synergistic effects of p-n heterojunction, supporting and zeolite nanoparticles in photo degradation of Eroochrome Black T and Methyl Orange,J . Photochem, Photobiol. A:chem. 346(2017) 133-143.
29. Pouretedal, H. R., Fallahgar, M., Pourhasan, F. S., & Nasiri, M. (2017). Taguchi optimization of photodegradation of yellow water of trinitrotoluene production catalyzed by nanoparticles TiO2/N under visible light. Iranian Journal of Catalysis, 7(4).
30. Buthiyappan, A., Raman, A. A. A., & Daud, W. M. A. W. (2016). Development of an advanced chemical oxidation wastewater treatment system for the batik industry in Malaysia. RSC advances, 6(30), 25222-25241.
31. Ajoudanian, N., & Nezamzadeh-Ejhieh, A. (2015). Enhanced photocatalytic activity of nickel oxide supported on clinoptilolite nanoparticles for the photodegradation of aqueous cephalexin. Materials Science in Semiconductor Processing, 36, 162-169.
32. Derikvandi, H., & Nezamzadeh-Ejhieh, A. (2017). Increased photocatalytic activity of NiO and ZnO in photodegradation of a model drug aqueous solution: effect of coupling, supporting, particles size and calcination temperature. Journal of hazardous materials, 321, 629-638.
33. Shabani, L., & Aliyan, H. (2024). Synthesis and photocatalytic activity of nanosized modified mesocellulous silica foams (MCFs) with PW12 and vanadium oxide. Iranian Journal of Catalysis, 6(3).
34. Corma, A., & Garcia, H. (2008). Supported gold nanoparticles as catalysts for organic reactions. Chemical Society Reviews, 37(9), 2096-2126.
35. Xu, L., Dai, Z., Duan, G., Guo, L., Wang, Y., Zhou, H., ... & Li, T. (2015). Micro/nano gas sensors: A new strategy towards in-situ wafer-level fabrication of high-performance gas sensing chips. Scientific reports, 5(1), 10507.
36. Swathi, D., Sabumon, P. C., & Maliyekkal, S. M. (2017). Microbial mediated anoxic nitrification-denitrification in the presence of nanoscale oxides of manganese. International Biodeterioration & Biodegradation, 119, 499-510.
37. Kumar, S., Chauhan, P., & Kundu, V. (2016). Sol–gel synthesis, structural, morphological and optical properties of Se-doped $SnO_2$ nanoparticles. Journal of Materials Science: Materials in Electronics, 27, 3103-3108.
38. Sharafzadeh, S., & Nezamzadeh-Ejhieh, A. (2015). Using of anionic adsorption property of a surfactant modified clinoptilolite nano-particles in modification of carbon paste electrode as effective ingredient for determination of anionic ascorbic acid species in presence of cationic dopamine species. Electrochimica Acta, 184, 371-380.
39. Fallah, N. S., & Mokhtary, M. (2015). Tin oxide nanoparticles ($SnO_2$-NPs): An efficient catalyst for the one-pot synthesis of highly substituted imidazole derivatives. Journal of Taibah University for Science, 9(4), 531-537.
40. Mei, L., Chen, Y., & Ma, J. (2014). Gas sensing of $SnO_2$ nanocrystals revisited: developing ultra-sensitive sensors for detecting the $H_2S$ leakage of biogas. Scientific reports, 4(1), 6028.
41. Huang, X., Wang, H., Niu, C., & Rogach, A. L. (2015). $SnO_2$ nanoarrays for energy storage and conversion. CrystEngComm, 17(30), 5593-5604.
42. Snaith, H. J., & Ducati, C. (2010). $SnO_2$-based dye-sensitized hybrid solar cells exhibiting near unity absorbed photon-to-electron conversion efficiency. Nano letters, 10(4), 1259-1265.
43. Stefik, M., Cornuz, M., Mathews, N., Hisatomi, T., Mhaisalkar, S., & Grätzel, M. (2012). Transparent, conducting Nb: $SnO_2$ for host–guest photoelectrochemistry. Nano letters, 12(10), 5431-5435.
44. Tammina, S. K., Mandal, B. K., Ranjan, S., & Dasgupta, N. (2017). Cytotoxicity study of Piper nigrum seed mediated synthesized $SnO_2$ nanoparticles towards colorectal (HCT116) and lung cancer (A549) cell lines. Journal of Photochemistry and Photobiology B: Biology, 166, 158-168.
45. Wu, S., Cao, H., Yin, S., Liu, X., & Zhang, X. (2009). Amino acid-assisted hydrothermal synthesis and photocatalysis of $SnO_2$ nanocrystals. The Journal of Physical Chemistry C, 113(41), 17893-17898.
46. Xing, L., Dong, Y., & Wu, X. (2018). $SnO_2$ nanoparticle photocatalysts for enhanced photocatalytic activities. Materials Research Express, 5(8), 085026.
47. Letifi, H., Litaiem, Y., Dridi, D., Ammar, S., & Chtourou, R. (2019). Enhanced Photocatalytic Activity of Vanadium-Doped $SnO_2$ Nanoparticles in Rhodamine B Degradation. Advances in Condensed Matter Physics, 2019(1), 2157428.

48. Palmer, G. B., & Poeppelmeier, K. R. (2002). Phase relations, transparency and conductivity in Ga2O3-SnO2-ZnO. Solid State Sciences, 4(3), 317-322.
49. Bolzan, A. A., Fong, C., Kennedy, B. J., & Howard, C. J. (1997). Structural studies of rutile-type metal dioxides. Structural Science, 53(3), 373-380.
50. Batzill, M., & Diebold, U. (2005). The surface and materials science of tin oxide. Progress in surface science, 79(2-4), 47-154.
51. Al-Hamdi, A. M., Rinner, U., & Sillanpää, M. (2017). Tin dioxide as a photocatalyst for water treatment: a review. Process Safety and Environmental Protection, 107, 190-205.
52. Rajeshwaran, P., Sivarajan, A., Raja, G., Madhan, D., & Rajkumar, P. (2016). Effect of tungsten (W 6+) metal ion dopant on structural, optical and photocatalytic activity of SnO 2 nanoparticles by a novel microwave method. Journal of Materials Science: Materials in Electronics, 27, 2419-2425.
53. Shabna, S., Dhas, S. S. J., & Biju, C. S. (2023). Potential progress in SnO2 nanostructures for enhancing photocatalytic degradation of organic pollutants. Catalysis Communications, 177, 106642.
54. A. Rahman, J. R. Jennings, A. L. Tan and M. M. Khan, ACS Omega, 2022, 7, 22089–22110 CrossRef CAS PubMed.
55. H. Wang, X. Li, X. Zhao, C. Li, X. Song, P. Zhang and P. Huo, Chin. J. Catal., 2022, 43, 178–214 CrossRef CAS.
56. P. P. Singh and V. Srivastava, RSC Adv., 2022, 12, 18245–18265 RSC.
57. Q. Xiao, Q. X. Tong and J. J. Zhong, Molecules, 2022, 27, 619 CrossRef CAS.
58. An overview into advantages and applications of conventional and unconventional hydro(solvo)thermal approaches for novel advanced materials design.M.C.M.D. de Conti a b c, S. Dey d, W.E. Pottker c, F.A. La Porta a b
59. Solvothermally synthesized oxygen-deficient SnO2 for the degradation of methyl orange dye under sunlight and LED light irradiation Rekha B Rajput, Rohidas B KaleResults in Chemistry 4, 100530, 2022
60. B. Yu, Y. Li, Y. Wang, H. Li, R. Zhang, Facile hydrothermal synthesis of SnO2 quantum dots with enhanced photocatalytic degradation activity: Role of surface modification with chloroacetic acid, J. Environ. Chem. Eng. 9 (4) (2021) 105618.
61. L. Yang, Y. Yang, T. Liu, X. Ma, S.W. Lee, Y. Wang, Oxygen vacancies confined in SnO2 nanoparticles for glorious photocatalytic activities from the UV, visible to near-infrared region, New J. Chem. 42 (2018) 15253–15262,
62. Hydrothermally engineered highly active Bi2WO6 photocatalyst for the degradation of Rhodamine B, Bisphenol A and dye mixture under visible light irradiation Gopika Rajan , Rekha B. Rajput, Rahilah S. Shaikh , Rohidas B. Kale *
63. Bhuvaneswari, K., Nguyen, B. S., Nguyen, V. H., Nguyen, V. Q., Nguyen, Q. H., Palanisamy, G., ... & Pazhanivel, T. (2020). Enhanced photocatalytic activity of ethylenediamine-assisted tin oxide (SnO2) nanorods for methylene blue dye degradation. Materials Letters, 276, 128173.
64. Xiao, H., Qu, F., Umar, A., & Wu, X. (2016). Facile synthesis of SnO2 hollow microspheres composed of nanoparticles and their remarkable photocatalytic performance. Materials Research Bulletin, 74, 284-290.
65. Viet, P. V., Thi, C. M., & Hieu, L. V. (2016). The high photocatalytic activity of SnO2 nanoparticles synthesized by hydrothermal method. Journal of Nanomaterials, 2016(1), 4231046.:-
66. Prakash, K., Senthil Kumar, P., Pandiaraj, S., Saravanakumar, K., & Karuthapandian, S. (2016). Controllable synthesis of SnO2 photocatalyst with superior photocatalytic activity for the degradation of methylene blue dye solution. Journal of Experimental Nanoscience, 11(14), 1138-1155.
67. Vinosel, V. M., Anand, S., Janifer, M. A., Pauline, S., Dhanavel, S., Praveena, P., & Stephen, A. (2019). Enhanced photocatalytic activity of Fe 3 O 4/SnO 2 magnetic nanocomposite for the degradation of organic dye. Journal of Materials Science: Materials in Electronics, 30, 9663-9677.

68. Suthakaran, S., Dhanapandian, S., Krishnakumar, N., & Ponpandian, N. (2019). Hydrothermal synthesis of SnO2 nanoparticles and its photocatalytic degradation of methyl violet and electrochemical performance. Materials Research Express, 6(8), 0850i3.